\documentclass[12pt, draftclsnofoot, onecolumn]{IEEEtran}

\usepackage{cite}
\usepackage{amsmath,amssymb,amsfonts}
\usepackage{algorithmic}
\usepackage{graphicx}
\usepackage{textcomp}
\usepackage{xcolor}
\usepackage{subfigure}
\usepackage{amsthm}
\usepackage{mathrsfs}

\usepackage{pdfpages}
\usepackage{colortbl}
\definecolor{mygray}{gray}{0.85}
\usepackage{array}
\usepackage{caption}
\usepackage{setspace}
\usepackage[linesnumbered,boxed,ruled,commentsnumbered]{algorithm2e}
\usepackage[ruled,linesnumbered]{algorithm2e}

{}
{}
{}

\newtheorem{case}{Case}{}

\usepackage{amsmath,bm}
\usepackage{cite}
\usepackage{amsmath,amssymb,amsfonts}
\usepackage{algorithmic}
\usepackage{amsthm}
\usepackage{graphicx}
\usepackage{textcomp}
\usepackage{xcolor}

\begin{document}
\begin{spacing}{1.5}
\title{HAP-Reserved Communications in Space-Air-Ground Integrated Networks}

\author{Xuelin~Cao,~
        Bo~Yang,~\IEEEmembership{Member,~IEEE,}
        Chau~Yuen,~\IEEEmembership{Fellow,~IEEE,}
        and~Zhu~Han,~\IEEEmembership{Fellow,~IEEE}

}
\maketitle

\begin{abstract}
Terrestrial communication networks have experienced significant development in recent years by providing emerging services for ground users. However, one critical challenge raised is to provide full coverage (especially in dense high-rise urban environments) for ground users due to scarce network resources and limited coverage. To meet this challenge, we propose a high altitude platform (HAP)-reserved ground-air-space (GAS) transmission scheme, which combines with the ground-to-space (G2S) transmission scheme to strengthen the terrestrial communication and save the transmission power. To integrate the two transmission schemes, we propose a transmission control strategy.  Wherein, the ground user decides its transmission scheme, i.e., switches between the GAS link transmission and the G2S link transmission with a probability. We then maximize the overall throughput and derive the optimal probability that a ground user adopts the GAS transmission scheme. Numerical results demonstrate the superiority of the proposed transmission control strategy.





\end{abstract}

\begin{IEEEkeywords}
    Space-air-ground integrated network (SAGIN), high altitude platform (HAP), reservation transmission. 
	\end{IEEEkeywords}

\IEEEpeerreviewmaketitle

\section{Introduction}
In recent years, traditional terrestrial communications have realized significant achievements in providing convenient communication services for ground users. However, due to economic or technological reasons, there is still a potential colossal requirement for the ground users that enjoy high-quality communication services from anywhere, such as rural, mountain, desert, and sea areas. Motivated by this challenge, the air and space communications have been leveraged to complement ground communications and provide more comprehensive coverage \cite{karapantazis2005broadband, Hassan, Zhang}. Compare to the ground communication system, the air and space communication systems have better coverage but suffer from limited capacity and long propagation latency. To overcome the mentioned shortcomings and implement the ground, air, and space communications preferably, some researchers have proposed the space-air-ground integrated network (SAGIN) to complement each other at different altitudes and achieve more flexible end-to-end services \cite{liu2018space, zhu2019cooperative, zhang2017software}. By integrating the ground, air, and space segments, the transmissions from the ground users to the destinations can be initiated via various paths. As a result, the different segments can provide differential quality transmissions to meet distinguished services, which is essential for the next-generation communication networks. Currently, extensive works have been investigated in the SAGIN, such as machine learning \cite{kato2019optimizing}, non-orthogonal multiple access (NOMA) \cite{zhu2017non}, edge computing \cite{zhang2019satellite}, reconfigurable intelligent surface (RIS) \cite{XCao}, and spatial resource allocation \cite{zhang2019spatial,du2017resource}. And even more recently, the related works on UAV-aided SAGIN have been explored deeply to improve system capacity and reduce power consumption \cite{JWang,YWang}. However, the research on the effective cooperation among ground, air, and space communications is still in its infant stage, especially for transmission access and control at the medium access control (MAC) layer of SAGIN.

\begin{figure}[t]
\centering{\includegraphics[width=3.5in,height=2.6in]{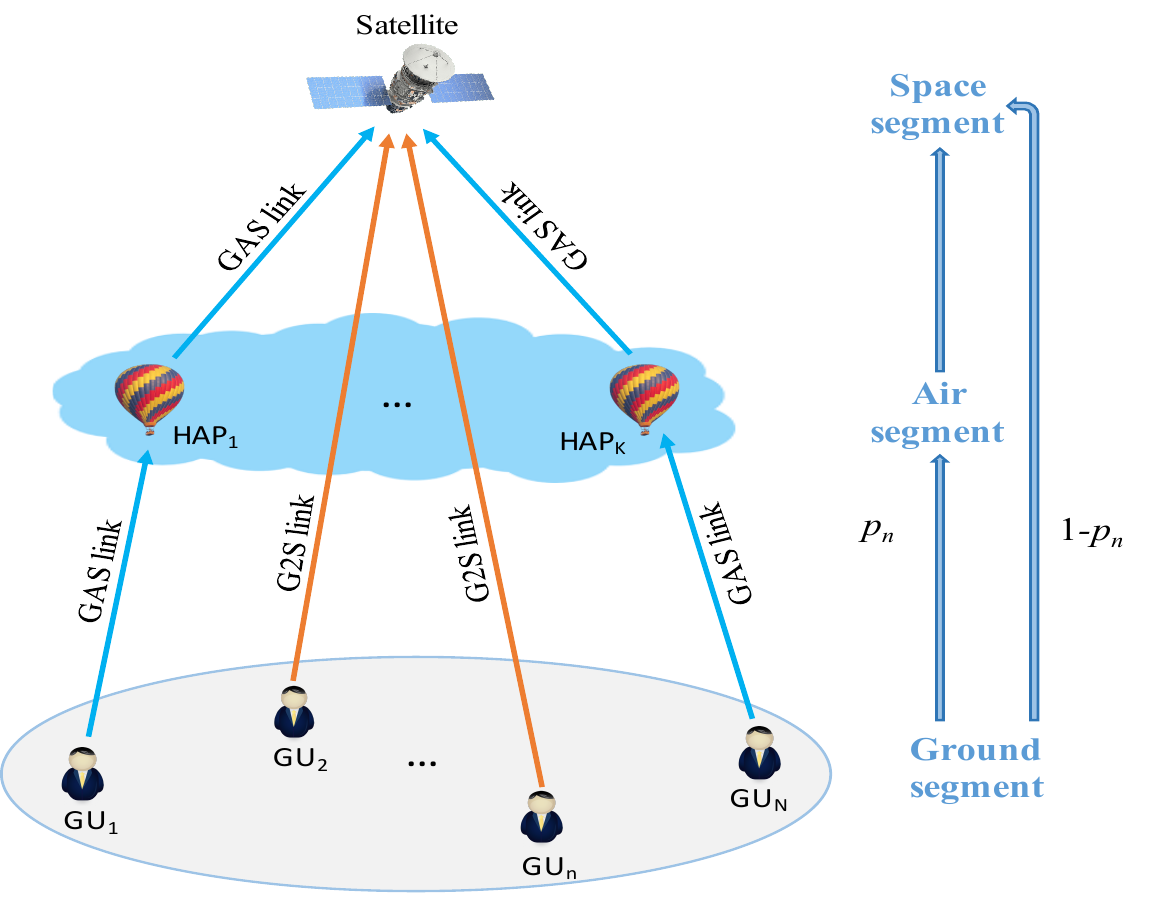}}
	\caption{\small A considered uplink ground-space communication system, where GAS links are proposed to complement G2S links. In particular, a transmission control strategy is presented, where the ground user $n$ initiates the GAS link transmission with probability $\rho_n$, while initiates the G2S link transmission with probability $1-\rho_n$.}
	\label{sys}
\end{figure}

In this paper, we mainly focus on integrating ground, air, and space communications by implementing a transmission control strategy, as highlighted in Fig. \ref{sys}. To complement the terrestrial communication system, we propose a high altitude platform (HAP)-reserved ground-air-space (GAS) transmission scheme to support ground-to-space (G2S) link transmissions, thereby achieving flexible and differentiated services, efficient spectrum utilization, energy-saving, and coverage enhancement. In the proposed GAS transmission scheme, HAPs can be reserved by the ground users to support their GAS link transmissions, where a GAS link includes a ground-to-air link and an air-to-space link. Based on the proposed GAS and G2S transmission schemes, we explore a transmission control strategy that enables ground users to adopt the GAS link transmission with a certain probability. Furthermore, we formulate an optimization problem to maximize the throughput, thereby deriving the optimal probability and optimal HAP for each ground user. Finally, numerical results demonstrate the advantage of the proposed transmission control strategy.

We arrange the rest of the paper as follows. In Section \ref{sec2}, we describe the system model. Then, we formulate a transmission strategy optimization problem and give the problem solution in Section \ref{sec3}. Section \ref{sec4} analyzes the overall throughput performance and presents numerical results. Section \ref{sec6} concludes the paper.

\section{System Model}\label{sec2}
The system is presented in Fig. \ref{sys}, where three-segment networks are constructed by the ground, air, and space systems. The ground system includes $N$ ground users, i.e., $\!{\cal N} \!=\! \{\!\text{GU}\!_1, \!\text{GU}\!_2, \!\ldots, \!\text{GU}\!_N\}$. The air system employs totally $K$ HAPs (e.g., airships or balloons) as carriers, i.e., $\mathcal{K}=\{\text{HAP}_1, \text{HAP}_2, \ldots, \text{HAP}_K\}$. As a relay, each HAP serves one ground user at a time. The space segment that comprises one narrowband satellite can connect with the ground and air segments. As shown in Fig. \ref{sys}, the HAPs deployed in the air segment provide the potential assistance for wireless communications between the ground users and the satellite. In the proposed system, the ground user decides to communicate with the satellite via the HAP or not, which means that the ground user $n, n\in \mathcal{N}$, may use the G2S link or the GAS link with the probability $\rho_n$ or $1-\rho_n$ accordingly. By performing negotiation, the ground user is aware of whether HAPs are available and selects the optimal HAP for its GAS link transmission if available, thereby impacting the system performance. Without loss of generality, we assume that the number of HAPs is limited, which cannot provide services for all ground users. In addition, the total system bandwidth is $B$, the bandwidth for the G2S links and the GAS links is $\omega_1B$ and $\omega_2B$, respectively. Here, $\omega_1$ and $\omega_2$ are the weights of the allocated bandwidth, and $\omega_1+\omega_2\le1$ holds.

We define $\rho_n, \rho_n\in[0,1]$, as the transmission probability of $\text{GU}_n$ via a HAP, and $1-\rho_n$ denotes the transmission probability of $\text{GU}_n$ without the assistance of HAPs. Specifically, we have
\begin{equation}
\centering 
 \!\rho_n =
\left\{\begin{array}{l}
 \!0,\;\;\;\;\;\;  \text{GU}_n\ \text{only use G2S link},  \\
 \!1,\;\;\;\;\;\;  \text{GU}_n\ \text{only use GAS link}, \\
 \!(0,1),  \text{GU}_n\ \text{uses GAS link with a probability}.
\end{array} \right.
\end{equation}
 
In addition, let $\mathbf{U}_n=\{{u}_{n1},{u}_{n2},\ldots,{u}_{nK}\}, \forall n \in \mathcal{N}$, denote the $K$-dimensional link decision vector of the $\text{GU}_n$, where ${u}_{nk} \in \{0, 1\}$ indicates the $\text{GU}_n$ communicates with the satellite via HAP $k$ or not. Then we have
\begin{equation}
\centering
u_{nk} =
\left\{\begin{array}{l}
1,\;\;\;\; \text{if}\ \text{GU}_n\ \text{transmits via the HAP}\ k,  \\
0,\;\;\;\; \text{otherwise}.
\end{array} \right.
\end{equation}

Moreover, we consider that $\text{GU}_n$ either transmits without the assistance of HAPs or transmits only via a HAP, then we have
\begin{equation}
\centering
\sum_{k=0}^{K} u_{nk} =
\left\{\begin{array}{l}
1,\;\;\;\; \text{if}\ \text{GU}_n\ \text{transmits via a HAP},  \\
0,\;\;\;\; \text{otherwise}.
\end{array} \right.
\end{equation}

Therefore, the transmission strategy of $\text{GU}_n$ is illustrated as
\begin{equation}
\centering
\mathcal{D}_n=\{\rho_n, \mathbf{U}_n\}. 
\end{equation} 

Based on the above illustrations, the transmission control decision for ground users can be given by $\mathbb{D}=\{\mathcal{D}_1,\ldots, \mathcal{D}_N\},\;\; \forall n \in \mathcal{N}$.  

\subsection{G2S Link Transmission Model}\label{sec2a}

\subsubsection{Channel Access Scheme}\label{sec2a1}
For the G2S link transmission, the ground users communicate with the satellite using an orthogonal frequency-division multiple access (OFDMA)-based carrier-sense multiple access/collision avoidance (CSMA/CA) scheme. Suppose that there are $ M $ available sub-frequency bands, i.e., $\mathcal{M}=\{1, 2, \ldots, M\}$, the bandwidth of each sub-frequency is $\omega_1B/M$. Based on CSMA/CA scheme, a ground user competes for the transmission access on $ M $ sub-frequency band and then transmits to the satellite via the G2S link. 

\subsubsection{Propagation Model}\label{sec2a1}
The G2S link transmission is a direct link transmission. According to the path loss propagation model, the G2S link rate for $\text{GU}_n$ is denoted by 
\begin{equation}\label{data rate}
\centering
R_n^{G2S}=\frac{\omega_1B}{M}\log_{2}\left(1+\frac{E_n^{G2S}10^{-L_n^{G2S}/10}}{\sigma_0^2}\right), 
\end{equation} 
where $E_n^{G2S}$ denotes the transmit power at the $\text{GU}_n$ using the G2S link, $L_n^{G2S}$ is the path loss from the $\text{GU}_n$ to the satellite, and $\sigma_0^2$ is Gaussian noise power of the G2S link.

\subsection{GAS Link Transmission Model}\label{sec2b}

\subsubsection{Channel Access Scheme}\label{sec2b1}
For the proposed GAS transmission scheme, the ground users communicate with the satellite using time-division multiple access (TDMA)-based reservation scheme, where the time is divided into a series of frames. Each frame consists of two periods: the ground negotiation period with the length of $t_h$ and the reserved GAS transmission period with the length of $t_r$ \cite{cao2019distributed}. Thus, the length of one time frame, denoted as $T$, is equal to $t_h+t_r$. During the ground negotiation period, following the specified licensed assisted access (LAA) listen-before-talk (LBT) scheme in 3GPP release 14, the ground users compete for the reservation privilege on the unlicensed band to reserve the HAP and slots. In 3GPP standard protocol, LAA defines the spectrum sharing on the unlicensed band. Specifically, LAA adopts a scheme called LBT, which allows the licensed users to leverage the unlicensed band when the channel is sensed to be idle, thus improving data rates and providing better services \cite{XC}. During the reserved GAS transmission period, the ground users communicate with the satellite via the reserved HAP within the reserved slots. Compared with the G2S link transmission, the proposed GAS transmission scheme can reduce the power consumption of the ground user.

\subsubsection{Propagation Model}\label{sec2b2}
The GAS link transmission can be considered as an amplify-and-forward transmission, each GAS link transmission occupies $\omega_2B$ bandwidths since only one GAS link transmission can be allowed to exit at a time. According to the path loss propagation model, the GAS link rate via the HAP$_k$ for $\text{GU}_n$ is denoted by 
\begin{equation}\label{data rate1}
\centering
R_{nk}^{GAS}=\omega_2B\log_{2}\left(1+f\left(r_{nk}, r_{ks}\right)\right), 
\end{equation} 
where 
\begin{equation}\label{eq1}
\centering
f\left(r_{nk}, r_{ks}\right)= \frac{r_{nk}r_{ks}}{r_{nk}+r_{ks}+1}. 
\end{equation}
In \eqref{eq1}, $r_{nk}$ and $r_{ks}$ denote the signal-to-noise ratio (SNR) at the HAP$_k$ and the satellite, respectively. They are expressed as 
\begin{equation}
\centering
r_{nk}= \frac{E_{n}^{GAS}10^{-L_{nk}^{GAS}/10}}{\sigma_1^2}, 
\end{equation}
and 
\begin{equation}
\centering
r_{ks}= \frac{E_{k}^{GAS}10^{-L_{ks}^{GAS}/10}}{\sigma_2^2}, 
\end{equation}
where $E_{n}^{GAS}$ and $E_{k}^{GAS}$ are the transmit power at the $\text{GU}_n$ and HAP$_k$, respectively. $L_{nk}^{GAS}$ and $L_{ks}^{GAS}$ are the path loss from $\text{GU}_n$ to HAP$_k$ and from HAP$_k$ to the satellite, respectively. $\sigma_{1}^2$ and $\sigma_{2}^2$ are Gaussian noise power from $\text{GU}_n$ to HAP$_k$ and from HAP$_k$ to the satellite, respectively.

\vspace{-5pt}
\section{Problem Formulation and Solution}\label{sec3}\subsection{Problem Formulation }\label{sec3a}
Having the defined system model and the proposed transmission control strategy in Section \ref{sec2}, we formulate the optimization problem to maximize the overall throughput of all ground users as follows:
\begin{align}\label{p1}
\mathscr{P}1: &\mathop {\max}\limits_{\mathbb{D}} \!\left(\!\sum_{m=1}^{M}\!\sum_{n=1}^{N}\!\alpha_n(1\!-\!\rho_n)\!R_n^{G2S}\!+\!\sum_{n=1}^{N}\!\sum_{k=1}^{K}\!\beta_n\rho_nu_{nk}\!R_{nk}^{GAS}\!\right)
\notag
\\
\bf{s.t.}\ \
&\text{C1:}\ 0 \leq \ \rho_n \leq 1, \ \ \ \ \ \ \ \ \ \ \ \ \ \ \ \ \ \forall n,
\notag\\
&\text{C2:}\ \ u_{nk}\in\{0,1\}, \ \ \ \ \ \ \ \ \ \ \ \ \ \ \ \ \forall n, \exists k,
\notag\\
&\text{C3:}\ \ \sum_{k=1}^{K}u_{nk}\in\{0,1\}, \ \ \ \ \ \ \ \ \ \ \ \forall n, \exists k,
\notag\\
&\text{C4:}\ \ \sum_{n=1}^{N}\sum_{k=1}^{K}u_{nk}= N_{s}, \ \ \ \ \ \ \ \ \ \ \forall n, \exists k,
\notag\\
&\text{C5:}\ \ 0<\alpha_n < 1, 0<\beta_{n} < 1, \ \ \ \forall n,
\end{align}
where $\alpha_n$ and $\beta_n$ are defined as the channel utilization of the G2S transmission scheme and the GAS transmission scheme, respectively. $\text{C1}$ limits the transmission probability of GAS link of $\text{GU}_n$. $\text{C2}$ and $\text{C3}$ are the constrains of $u_{nk}$. $\text{C4}$ constraints the number of ground users that adopt the GAS link transmission. $N_{s}$ is the number of ground users that successfully reserve HAP, which is affected by the competition process during the period of $t_h$. $\text{C5}$ shows the feasibility constraints of $\alpha_n$ and $\beta_n$.

In problem $\mathscr{P}1$, the overall throughput includes the throughput of G2S and GAS links. Let $N_1$ be the number of ground users that use the OFDMA-based G2S link transmission to connect the satellite. Refer to \cite{bianchi2000performance}, as $N_1$ ground users compete for access at a sub-channel, the successful transmission probability ($p_s$), the idle probability ($p_e$), and the failed transmission probability ($p_c$) can be expressed as
\begin{equation}\label{q1}
\centering
\left\{\begin{array}{l}
p_s\!=\!N_1\tau(1\!-\!\tau)^{\!N_1\!-\!1}, \\
p_e\!=\!(1\!-\!\tau)^{N_1},\\
p_c\!=\!1\!-\!p_s\!-\!p_e
\end{array} \right.
\end{equation}
where $\tau$ is the stationary probability that a ground user transmits in a random slot, which is denoted by
\begin{equation}\label{q2}
\centering
\tau\!=\!\frac{2(1-2p)}{(1 - 2p)(W+1) + pW(1 - {(2p)}^l)}.
\end{equation}
In \eqref{q2}, $W\in[W_{min},W_{max}]$ is the contention window and $l$ is the backoff stage. $W_{min}$ and $W_{max}$ denote the minimum and maximum contention window, respectively. $p$ is the collision probability, which is expressed as  
\begin{equation}\label{q3}
\centering
p\!=\!1\!-(1\!-\!\tau)^{\!N_1\!-\!1}.
\end{equation}
According to \cite{bianchi2000performance}, we can obtain the channel utilization of each sub-channel as 
\begin{equation}\label{q4}
\centering
\varphi=\frac{p_st_s}{p_e\delta+p_st_s+p_ct_c},
\end{equation}
where $t_s\!=\!RTS\!+CT\!S\!+\!t_{p}\!+\!2{SIFS}\!+\!{DIFS}\!+\!2\delta$ and $t_c\!=\!RTS\!+\!DIFS\!+\!\delta$ are the time length that the channel undergoes a successful transmission and failed transmission, respectively. $t_p$ is the time length of a payload. $RTS$, $CTS$, $SIFS$, and $DFIS$ are the duration of request-to-send (RTS), clear-to-send (CTS), short inter-frame space (SISF), and DCF inter-frame space (DISF), respectively.

For the proposed GAS transmission scheme, since $\text{GU}_n$ can use the G2S link transmission on $M$ sub-channels with probability $1-\rho_n$, the total throughput of GAS link transmissions is denoted by
\begin{equation}\label{G2S}
\centering
\mathbb{S}^{G2S}=\sum_{m=1}^{M}\sum_{n=1}^{N}\frac{p_st_s}{p_e\delta+p_st_s+p_ct_c}(1-\rho_n)R_n^{G2S}.
\end{equation} 

Furthermore, when $N_2$ ground users adopt the TDMA-based GAS transmission scheme to connect with the satellite, the total throughput of GAS links within a frame is expressed as 
\begin{align}\label{GAS}
\mathbb{S}^{GAS}&=\sum_{n=1}^{N}\rho_nS_{nk}^{GAS}, 
\end{align}
where $S_{nk}^{GAS}$ denotes the throughput of a ground user that adopts the GAS link transmission, which is denoted by
\begin{align}\label{thrj}
\begin{split}
S_{nk}^{GAS}&=\sum_{k=1}^{K}\frac{t_pr_n}{T}u_{nk}R_{nk}^{GAS}. 
\end{split}
\end{align}
In \eqref{thrj}, $r_n$ denotes the reservation step, i.e., the number of data packets that $\text{GU}_n$ can transmit in one frame. $\!\frac{t_pr_n}{T}$ is the channel utilization when $\text{GU}_n$ adopts the GAS link transmission. Due to the limitation of the ground negotiation period and the reserved GAS transmission period, the number of successful reserved ground users should meet the condition that $\sum_{n=1}^{N}\sum_{k=1}^{K}u_{nk}\!=\!N_s\!=\!\frac{t_h\zeta_s}{t_s'}$. Here, $t_s'\!=\!{RTS}\!+\!{CTS}\!+\!{SIFS}\!+\!{DIFS}$ is the required negotiation time for $\text{GU}_n$, $\zeta_s$ is the probability of successful negotiation for $\text{GU}_n$. Refer to \cite{cao2019distributed}, $\zeta_s$ is given by
\begin{align}\label{GAS1}
\zeta_s = N_2q\epsilon(1-\epsilon)^{N_2q-1}, 
\end{align}
where $\epsilon = \frac{2(1-2\varrho)q}{{q[(W + 1)(1 - 2\varrho) + pW(1 - {{(2\varrho)}^l})] + 2(1 - q)(1 - \varrho)(1 - 2\varrho)}}$  represents the stationary probability that $\text{GU}_n$ transmits a data packet in a random slot during the ground negotiation period. $\varrho=1-(1-\epsilon)^{N_2q-1}$ is the collision probability of $\text{GU}_n$. $q = \frac{\eta }{\gamma  + \eta}$ is the stationary probability that $\text{GU}_n$ stays in the ground negotiation period. Here, $\gamma$ and $\eta$ denote the arrival rate and the service rate, respectively.

Then, the total throughput $\mathbb{S}^{GAS}$ can be expressed as 
\begin{align}\label{GAST}
\mathbb{S}^{GAS}&=\sum_{n=1}^{N}\sum_{k=1}^{K}\frac{t_pr_n}{T}\rho_nu_{nk}R_{nk}^{GAS}. 
\end{align}

\subsection{Problem Solution}\label{sec3b}

According to the above throughput analysis in \eqref{G2S} and \eqref{GAST} for each scheme, we have $\alpha_n=\frac{p_st_s}{p_e\delta+p_st_s+p_ct_c}$ and $\beta_n=\frac{t_pr_n}{T}$ for $\text{GU}_n$, where  $p_s\!=\!N_1\tau(1\!-\!\tau)^{N_1-1}$, $p_e\!=\!(1\!-\!\tau)^{N_1}$, $p_c\!=\!1\!-\!p_s\!-\!p_e$, $r_n\!=\!\frac{t_r}{N_st_p}\!=\!\frac{t_rt_s'}{t_pt_h\zeta_s}$, and $\zeta_s = N_2q\epsilon(1-\epsilon)^{N_2q-1}$. Thus, the optimization problem $\!\mathscr{P}1$ is rewritten as problem $\!\mathscr{P}2$, which is given by 
\begin{align}\label{p2}
\mathscr{P}2:&\mathop {\max}\limits_{\mathbb{D}} \left(\sum_{m=1}^{M}\sum_{n=1}^{N}\frac{p_st_s}{p_e\delta+p_st_s+p_ct_c}(1-\rho_n)R_n^{G2S}
\notag\right.
\\
\phantom{=\;\;}
&\left.+\sum_{n=1}^{N}\!\frac{t_pr_n}{\!T}\rho_n\sum_{k=1}^{K}u_{nk}R_{nk}^{GAS}\right) 
\notag\\
\bf{s.t.}\ \
&\text{C1-C5},\notag \\
&\text{C6:}\ \ N_1=N-\sum_{n=1}^{N}\rho_n,\ N_2=\sum_{n=1}^{N}\rho_n,\ \ \forall n.
\end{align}

To simplify problem $\mathscr{P}2$, we assume that $\rho_n=\rho, r_n=r, \forall n\in\mathcal{N}$, then we have $N_1=N(1-\rho)$ and $N_2=N\rho$. Since $p_s$, $p_e$, and $p_c$ are related to the value of $N_1$, and $\zeta_s$ is related to the value of $N_2$, it is observed that $\alpha_n$ and $\beta_n$ are the functions of $\rho$. Let $\alpha_n=\alpha(\rho)$ and $\beta_n=\beta(\rho)$, then we have 
\begin{align}\label{p3}
\mathscr{P}3:&\mathop {\max}\limits_{\mathbb{D}}\!\left(\!\alpha(\rho)\!(1\!-\!\rho)\!\sum_{m=1}^{M}\!\sum_{n=1}^{N}\!R_n^{G2S}\!+\!\beta(\!\rho)\!\rho\!\sum_{n=1}^{N}\!\sum_{k=1}^{K}\!u_{nk}\!R_{\!nk}^{\!GAS}\!\right) 
\notag\\
\bf{s.t.}\
&\text{C2-C4}, 
\notag\\
&\text{C7}, \ 0 \leq \ \rho \leq 1, \ \forall n,
\notag\\
&\text{C8}, \ 0 < \ \alpha(\rho) < 1, \ 0 < \ \beta(\rho) < 1, \ \forall n,
\notag\\
&\text{C9}, \ N_2=N\rho,\ N_1=N-N\rho.
\end{align}

Problem $\!\mathscr{P}3$ is a mixed-integer nonlinear programming (MINLP) problem. To solve this MINLP problem, $\!\mathscr{P}3$ can be decomposed into two sub-problems, i.e., the GAS link transmission probability sub-problem and the HAP selection sub-problem. These two sub-problems can be optimized iteratively in an alternating manner with one being fixed in each iteration until both reach the convergence.

For a given HAP selection, $\!\{\!\mathbf{U}_1, \!\ldots, \!\mathbf{U}_N\}$, problem $\mathscr{P}3$ can be transformed as
\begin{align}\label{p3-1}
\!\mathscr{P}3.1\!:\ &\!\mathop{\max}\limits_{\rho} \!\left(\!M\!\alpha(\rho)\!(1\!-\!\rho)\!\sum_{n=1}^{N}\!R_n^{G2S}\!+\!\beta(\!\rho)\!\rho\!\sum_{n=1}^{N}\!\sum_{k=1}^{K}\!u_{nk}\!R_{nk}^{GAS}\!\right)\!
\notag\\ 
\bf{s.t.}\ \ \ \ &\text{C7-C9}.
\end{align}
$\!\mathscr{P}3.1$ is convex, the optimal $\rho$ can be solved by the existed math tools.

Then, given the GAS link transmission probability, $\rho$, for the ground users, problem $\mathscr{P}3$ can be transformed as
\begin{align}\label{p3-2}
\mathscr{P}3.2:\ &\mathop {\max}\limits_{\!\{\!\mathbf{U}_1, \!\ldots, \!\mathbf{U}_N\}} \!\beta(\rho)\rho\!\sum_{n=1}^{N}\!\sum_{k=1}^{K}\!u_{nk}\!R_{nk}^{GAS} 
\notag\\
\bf{s.t.}\ \ \ \
&\text{C2-C4}.
\end{align}
Problem $\!\mathscr{P}3.2$ is integer linear programming, the optimal $\{\mathbf{U}_1, \!\ldots, \!\mathbf{U}_N\}$ for $\text{GU}_n$ can be solved by using the exhaustive method. The computing complexity is $\mathcal{O}(2^{NK})$ when the exhaustive method is adopted to solve the HAP selection sub-problem. On this basis, it is observed that the exhaustive method can be adaptive to the scenario with only a few number of ground users and HAPs. When massive ground users and HAPs exist in the networks, the complexity of the exhaustive method becomes very large since it exponentially increases with the number of ground users and HAPs. Besides, an optimal solution can be obtained through the alternating iteration method only when the number of iterations is large enough. To address these challenges, we can use the deep learning-based method to solve the HAP selection problem more efficiently \cite{yang2019computation}.

According to the optimization solution of $\mathbb{D}$, the implementation algorithm of the GAS transmission scheme is presented in Alg. \ref{algo}, where each ground user makes a transmission decision to communicate with the satellite via the GAS link or G2S link, and then reserves the HAP$_k$ to assist its GAS link transmissions if the GAS transmission scheme is selected.

\begin{algorithm}[t]
    \caption{HAP-reserved Algorithm}
    \label{algo}
    \textbf{Input:} $\mathbb{D}$\;
    \For{n= \rm 1 \rm to N}{
    \If{$\rho_n=0$} 
    { {\rm  GU}$_n$ transmits data to the satellite with G2S link\; 
       }   
    \ElseIf {$\rho_n=1$} 
              {    \For{k= \rm 1 \rm to K}
                 {
                 \If{$u_{nk}=1$}
                 {
                 {\rm GU}$_n$ reserves HAP$_k$ in the period of $t_h$\; 
                 }  
                 }
                 {\rm GU}$_n$ transmits data to the satellite via HAP$_k$ with GAS link in the period of $t_r$ \;  
              } 
    \ElseIf {$0<\rho_n<1$} 
              {{\rm GU}$_n$ uses G2S link with probability $\!1\!-\!\rho_n$\;
               {\rm GU}$_n$ uses GAS link with probability $\rho_n$\; 
               Repeats \textit{line} 7 to \textit{line} 12;
              } 
  }             
\end{algorithm}

\section{Performance Analysis and Evaluation}\label{sec4}
\subsection{Performance Analysis}\label{sec3b}
We present three cases to show the overall throughput of the ground system when ground users adopt GAS link with probability, $\rho_n$, i.e., $0<\rho_n<1$, $\rho_n=0$, and $\rho_n=1$.
\begin{case}\label{l1}
Given $0<\rho_n<1,\forall n\in\mathcal{N}$, $N_1=N-\sum_{n=1}^{N}\rho_n$, and $N_2=\sum_{n=1}^{N}\rho_n$, the overall throughput that ground users communicate with the satellite via HAP with probability $\rho_n$, $\mathbb{S}^{sum}$, is expressed as
\begin{align}\label{G2AS}
\mathbb{S}^{sum}&=\mathbb{S}^{G2S}+\mathbb{S}^{GAS}\\
\notag
\notag
&=\omega_1\!B\!\sum_{n=1}^{N}\!\frac{t_sp_s\!(1\!-\!\rho_n)\log_{2}\left(1\!+\!\frac{\!E_n^{G2S}10^{\!-L_n^{G2S}/10}}{\sigma^2}\right)}{\!p_e\delta\!+\!p_st_s\!+\!p_ct_c}\\
\notag
&+\!\omega_2\!B\!\sum_{n=1}^{N}\!\frac{t_rt_s'}{Tt_h\zeta_s}\rho_n\!\sum_{k=1}^{K}u_{nk}\log_{2}\!\left(1\!+\!f\!\left(r_{nk},r_{ks}\!\right)\right).
\end{align}
\end{case}

\begin{proof}
When $\rho_n\in(0,1), \forall n\in\mathcal{N}$, it means that $\text{GU}_n$ transmits via GAS link with a certain probability. In other words, $N_1$ ground users adopt the OFDMA-based G2S link transmission, and $N_2$ ground users adopt the GAS link transmission. Based on \eqref{data rate} and \eqref{data rate1}, by using $N_1=N-\sum_{n=1}^{N}\rho_n$ in \eqref{G2S}, and by using $N_2=\sum_{n=1}^{N}\rho_n$ and $r_n=\frac{t_r}{N_st_p}$ in \eqref{GAST}, we can obtain the overall throughput as shown in \eqref{G2AS}. 
\end{proof}

\begin{case}\label{l2}
Given $\rho_n=0,\forall n\in\mathcal{N}$, the overall throughput that ground users communicate with the satellite without the assistance of HAP, $\mathbb{S}^{G2S}$, is expressed as
\begin{align}\label{rate}
\centering
\mathbb{S}^{G2S}&=\sum_{m=1}^{M}\frac{p_s't_s}{p_e'\delta+p_s't_s+p_c't_c}\sum_{n=1}^{N}
R_n^{G2S}\\
\notag
&\!=\!\frac{\omega_1Bp_s't_s}{p_e'\delta\!+\!p_s't_s\!+\!p_c't_c}\sum_{n=1}^{N}\log_{2}\!\left(\!1\!+\!\frac{E_n^{G2S}10^{\!-L_n^{G2S}/10}}{\sigma^2}\!\right)\!.
\end{align} 
\end{case}

\begin{proof}
When $\rho_n=0, \forall n\in\mathcal{N}$, it means that all the ground users communicate with the satellite by adopting the G2S link transmission, i.e., $N_1=N$, and $N_2=0$. According to \eqref{G2S}, the overall throughput of all ground users can be calculated as in \eqref{rate}, where $p_s', p_e', p_c', \tau'$, and $p'$ can be calculated using $N$ instead of $N_1$ in \eqref{q1}-\eqref{q3}.
\end{proof}

\begin{case}\label{l3}
Given $\rho_n=1,\forall n\in\mathcal{N}$, the overall throughput that ground users communicate with the satellite via HAP, $\mathbb{S}^{GAS}$, is expressed as
\begin{align}\label{thr1}
\mathbb{S}^{GAS}&=\sum_{n=1}^{N}\frac{t_pr_n}{T}\sum_{k=1}^{K}u_{nk}R_{nk}^{GAS}\\
\notag
&=\!\frac{\omega_2Bt_rt_s'}{Tt_h\zeta_s'}\!\sum_{n=1}^{N}\!\sum_{k=1}^{K}\!u_{nk}\!\log_{2}\left(1\!+\!f\!\left(r_{nk},r_{ks}\!\right)\right). 
\end{align}
\end{case}

\begin{proof}
When $\rho_n=1, \forall n\in\mathcal{N}$, it means that all the ground users communicate with the satellite by adopting the GAS link transmission, i.e., $N_1=0$, and $N_2=N$. According to \eqref{GAS}, the overall throughput of all ground users can be calculated as in \eqref{thr1}, where $\zeta_s'$, $\varrho'$, $q'$, and $\epsilon'$ can be calculated using $N$ instead of $N_2$ in \eqref{GAS1}.
\end{proof}

Based on the analysis of three cases, we conclude that the performance of the GAS transmission scheme is better than the performance of the G2S transmission scheme when the number of ground users is small, and there is an existing rational range of $\rho$ to achieve the better system performance.

\begin{figure}[t]
\centering{\includegraphics[width=3.2in,height=2.4in]{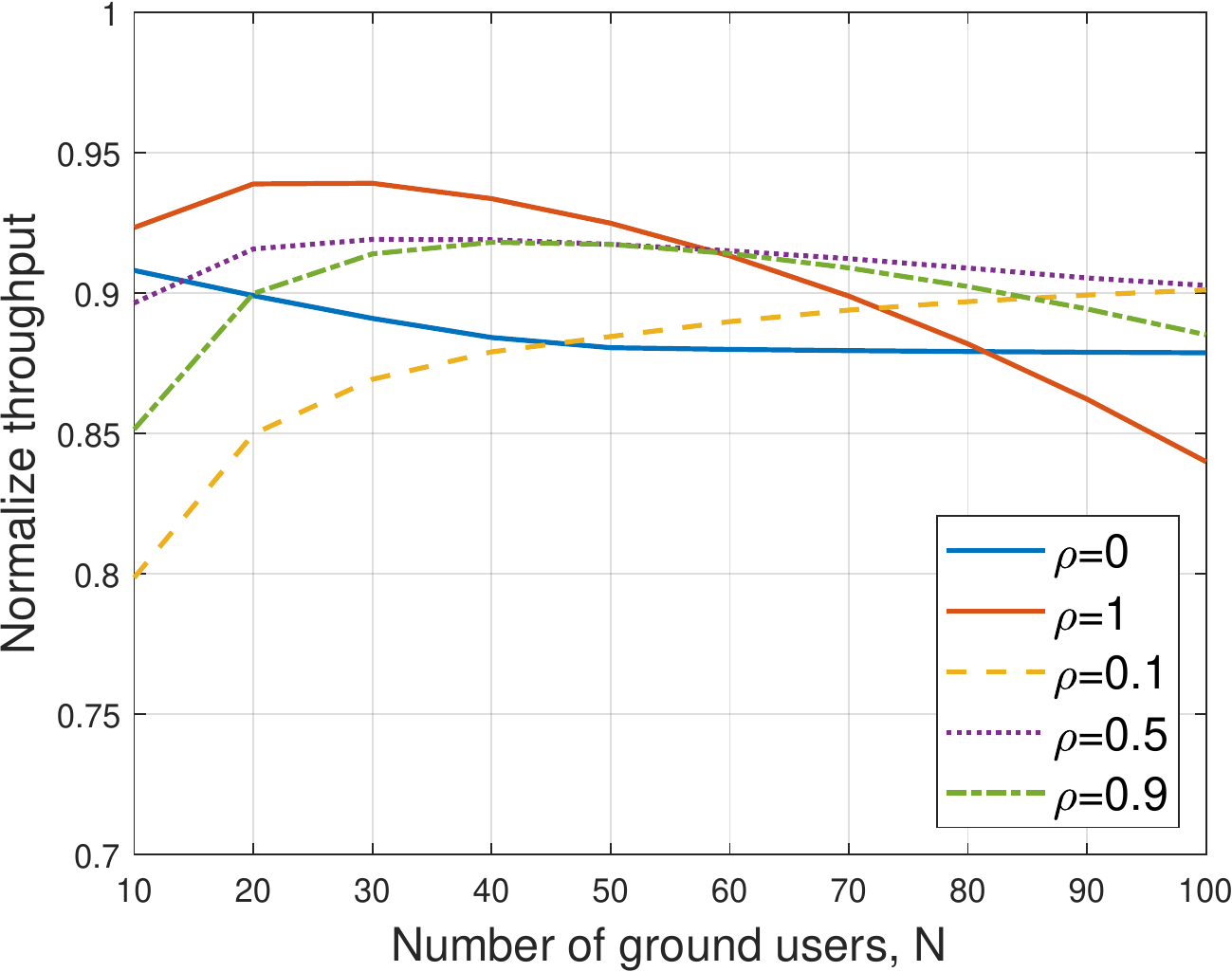}}
	\caption{\small Normalized throughput versus $N$.}
	\label{R2}
\end{figure}

\begin{figure}[t]
\centering{\includegraphics[width=3.1in,height=2.4in]{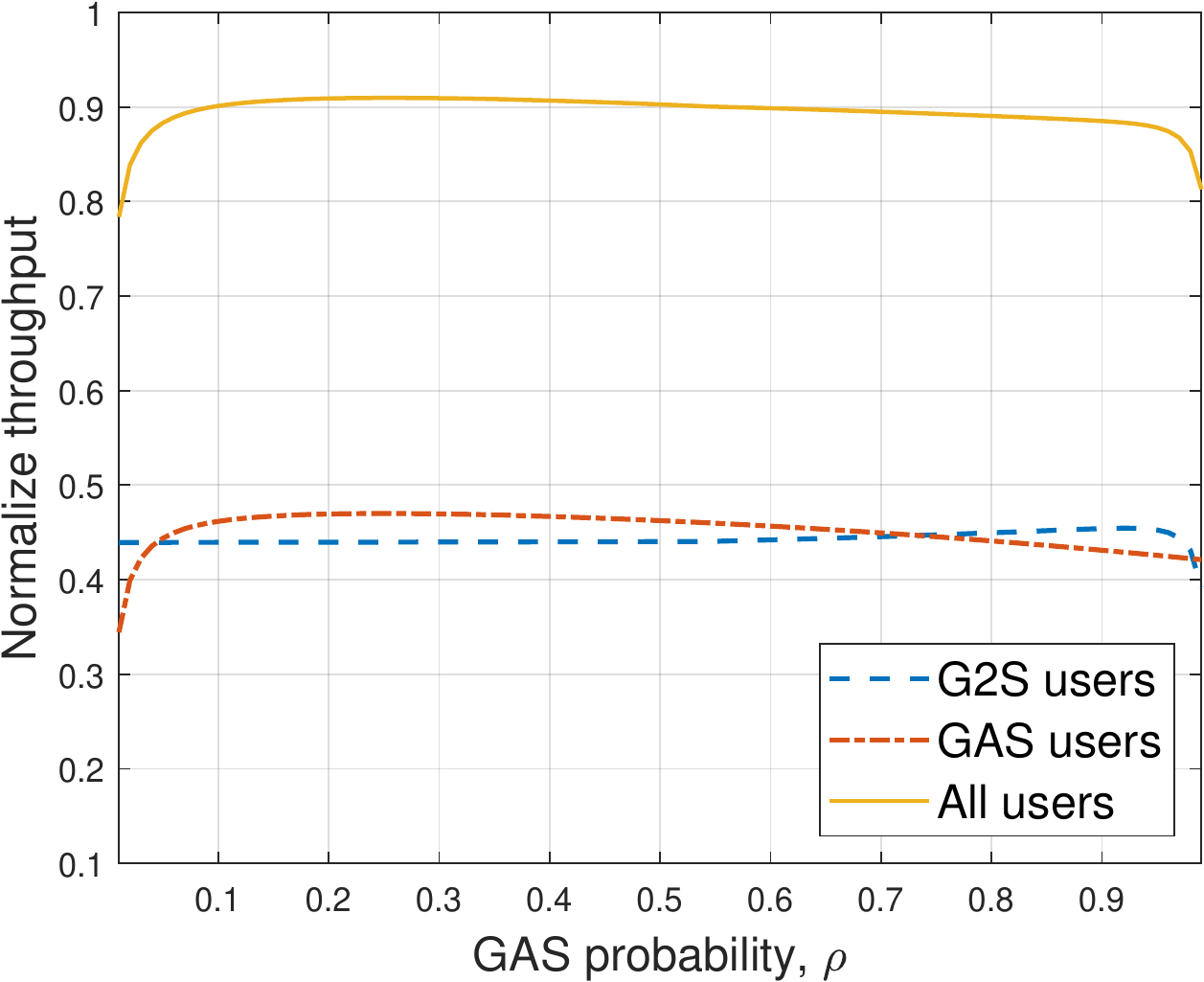}}
	\caption{\small Normalized throughput versus $\rho$.}
	\label{R1}
\end{figure}

\begin{figure}[t]
\centering{\includegraphics[width=3.2in,height=2.4in]{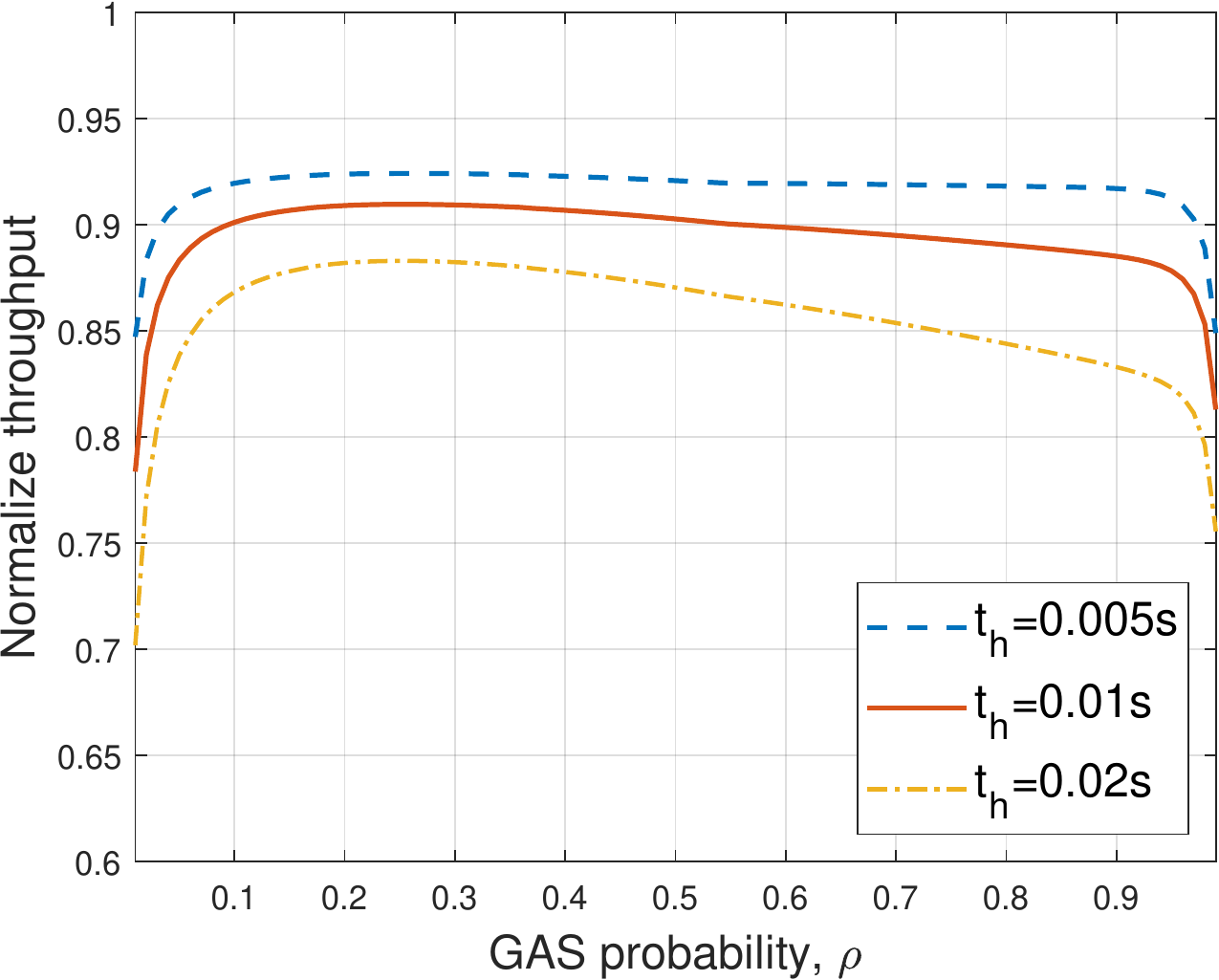}}
	\caption{\small Normalized throughput versus $\rho$.}
	\label{R3}
\end{figure}

\begin{figure}[t]
\centering{\includegraphics[width=3.2in,height=2.4in]{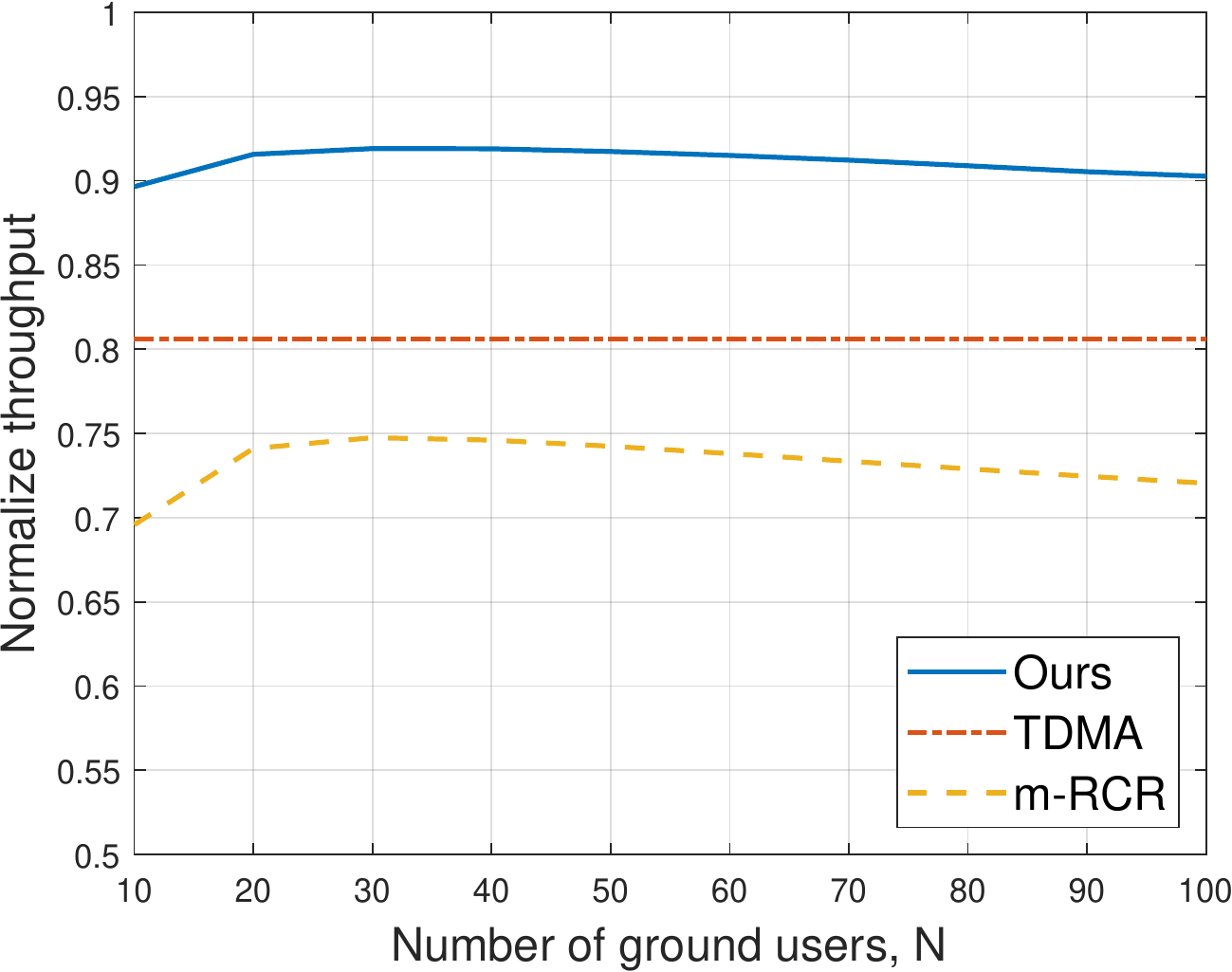}}
\caption{\small Normalized throughput versus $N$.}
\label{R4}
\end{figure}

\subsection{Performance Evaluation}\label{sec4a}	

In the following, we evaluate the performance of the proposed transmission control strategy. The key parameters are valued as follows: $N=100, M=5$, $T=200$ ms, $t_h=10$ ms, and $t_p=0.5$ ms. The packet size of RTS and CTS are $24$ bytes and $16$ bytes, respectively.

Figure \ref{R2} shows that the normalized throughput changes with the number of ground users ($N$) under the different GAS link probability ($\rho$). When $\rho=0$, the ground users adopt G2S link to transmit, the throughput of ground communication has a slight decline due to the competition collision. When $\rho=1$, each ground user selects a GAS link to communicate with the satellite. The throughput of ground communication first increases and then decreases. When $0<\rho<1$, each ground user uses a GAS link with probability $ \rho $ and uses a G2S link with probability $1-\rho$. We can observe from Fig. \ref{R2} that when $\rho$ is $0.1, 0.5$, and $0.9$, the throughput of ground communications in three cases have a similar trend, i.e., first increases and then declines. Therefore, the proposed transmission control strategy can balance G2S and GAS. Fig. \ref{R1} shows that the normalized throughput changes with the GAS link probability ($\rho$), where $0<\rho<1$. As shown in Fig. \ref{R1}, the throughput of G2S users and GAS users first climbs and then decreases as $\rho$ increases. Compare to the throughput of G2S users, the throughput of GAS users reaches the top earlier due to the low competition for GAS links and the high competition for G2S links when $\rho$ is low. 

Figure \ref{R3} shows that the normalized throughput changes with the probability ($\rho$) under the different ground negotiation periods ($t_h$). It can be seen that an optimal $\rho$ exists in three cases to maximize the system throughput. Besides, it also can be observed that a lower $t_h$ leads to a better throughput due to the low negotiation overhead. Fig. \ref{R4} shows that the normalized throughput changes with the number of ground users ($N$) in different schemes. Compared with the existed TDMA scheme and m-RCR scheme proposed in \cite {yang2019performance}, the proposed strategy that can be adaptive to the GAS and G2S link outperforms the other two schemes since the spectrum efficiency is improved. 

\section{Conclusion}\label{sec6}
In this paper, the HAP-reserved GAS transmission scheme is proposed to complement G2S link transmissions and strengthen conventional terrestrial communications. We investigate a transmission control strategy and explore a rational range of the GAS link transmission probability to maximize the system throughput. The numerical results show that the GAS transmission scheme can beat the G2S transmission scheme when the number of ground users is lower than 80, and achieve a better system performance when $\rho$ is in the range of $[0.1, 0.7]$. 


\end{spacing}
\end{document}